\documentclass[12pt]{article}

\newcommand{\bde}{\begin{description}}

\newcommand{\ben}{\begin{enumerate}}
\newcommand{\beq}{\begin{eqnarray}}

\newcommand{\beqn}{\begin{eqnarray*}}

\newcommand{\bqu}{\begin{quote}}

\newcommand{\bvb}{\begin{verbatim}}

\newcommand{\ea}{\end{eqnarray}}

\newcommand{\ede}{\end{description}}
\newcommand{\een}{\end{enumerate}}
\newcommand{\eeq}{\end{eqnarray}}
\newcommand{\eeqn}{\end{eqnarray*}}

\newcommand{\equ}{\end{quote}}

\newcommand{\evb}{\end{verbatim}}

\newcommand{\suppress}[1]{}
\newcommand{\supress}[1]{}

\newcommand{\pmbeg}{\begin{pmatrix}}
\newcommand{\pmend}{\end{pmatrix}}

\usepackage{amsmath,longtable}
\usepackage{rotating}
\usepackage{graphicx}
\usepackage{scicite, times}

\DeclareGraphicsExtensions{.eps, .jpg} 

\newenvironment{sciabstract}{
\begin{quote} \bf}
{\end{quote}}

\newcounter{lastnote}
\newenvironment{scilastnote}{
\setcounter{lastnote}{\value{enumiv}}
\addtocounter{lastnote}{+1}
\begin{list}
{\arabic{lastnote}.}
{\setlength{\leftmargin}{.22in}}
{\setlength{\labelsep}{.5em}}}
{\end{list}}

\title{ Inferring Network Mechanisms: The {\em Drosophila melanogaster} Protein Interaction Network}
\author{
Manuel Middendorf,$^{1}$
Etay Ziv,$^{2}$
Chris Wiggins,$^{3,4}$\\
\normalsize{$^1$Department of Physics,
$^2$College of Physicians and Surgeons,}\\
\normalsize{
$^3$Department of Applied Physics and Applied Mathematics,}\\
\normalsize{
$^4$Center for Computational Biology and Bioinformatics;}\\
\normalsize{
Columbia University, New York, NY 10027, USA}
}

\date{}

\begin{document}

\baselineskip24pt

\maketitle
\begin{sciabstract}
Naturally occurring networks exhibit 
quantitative 
features revealing underlying growth mechanisms. 
Numerous network mechanisms have 
recently
been proposed to reproduce 
specific 
properties such as degree distributions or clustering coefficients.
We present a method for inferring the mechanism most accurately capturing a given network topology, exploiting discriminative tools from machine learning.
The {\em Drosophila melanogaster} protein network is confidently and robustly 
(to noise and training data subsampling) 
classified as a duplication-mutation-complementation network 
over preferential attachment, small-world, and other duplication-mutation mechanisms.
Systematic classification, rather than statistical study of specific properties, provides a discriminative approach to understand the design of complex networks.

\end{sciabstract}

\section{Introduction}
Recent advances in our understanding of biological networks
 have often focused on understanding the emergence of specific
 features such as scale-free degree-distributions
\cite{global:strogatz:nature,Newman2003,barabasi},
short mean geodesic lengths or clustering coefficents \cite{watts}. 
The insights gained into the topological patterns have 
motivated various network growth and evolution models 
in order to determine what simple mechanisms can reproduce 
the features observed.
Among these are the preferential attachment model \cite{barabasi,Price65} 
exhibiting scale-free degree distributions, and  the small-world 
model \cite{watts} exhibiting  high clustering coefficients and 
short mean geodesics. 
Moreover, various duplication-mutation mechanisms have been 
proposed to describe biological networks 
\cite{flammini,Sole,wagner,model:rzethsky:bioinf,model:gerstein:jmb,model:dewey:bioinf} 
and the World Wide Web \cite{kumar}. 
However, in most cases model parameters can be tuned such that 
multiple models of widely varying mechanisms perfectly fit the 
motivating real network in terms of single selected features 
such as the scale-free exponent and the clustering coefficient.
Since networks with several thousands of vertices and edges 
are highly complex, it is also clear that these features can 
only capture limited structural information.  

Here, we make use of {\it discriminative
classification} techniques recently 
developed in  machine learning \cite{hastie,devroye} to 
classify a given real network as one of many proposed 
network mechanisms by enumerating local substructures.
Determining what simple mechanism is responsible for a 
natural network's architecture would 
(i) facilitate the development of correct priors for constraining 
network inference and reverse engineering
\cite{saito,goldberg,quaid,gomez}; 
(ii) specify the appropriate null model relative to 
which one evaluates statistical significance 
\cite{motifs:alon:nat,motifs:alon:science,motifs:hasty:nature,motifs:lee:science,motifs:barabasi:natgen,motifs:vespignani:natgen,motifs:alon:jmb,motifs:alon:pnas,matstat};
(iii) guide the development of improved network models; and 
(iv) reveal underlying design principles of evolved biological networks.
It is therefore desirable to develop a method 
to determine which proposed mechanism models a given complex network without 
prior feature selection.

Enumeration of subgraphs has been succesfully used
to find network motifs 
\cite{motifs:alon:nat,motifs:alon:science,motifs:hasty:nature,motifs:lee:science,motifs:barabasi:natgen,motifs:vespignani:natgen,motifs:alon:jmb,motifs:alon:pnas,matstat} 
during the past few years and is historically a well
established method in the sociology community \cite{holland}.
Recently, the idea of clustering real networks based on their ``significance
profiles'' has been proposed \cite{alon:cluster}. The method assumes
randomized networks with fixed degree distribution as the null model to 
estimate the statistical significance of given subgraphs. The significance 
profiles are then shown to be similar for various groups of naturally occurring
networks.  

Finding statistically significant motifs and clustering can both be 
characterized as schemes to identify a reduced-complexity description
of the networks.
We here present an approach which is instead {\em predictive},
in which labeled graphs of known growth mechanisms are used as training
data for a  discriminative classifier. This classifier, then, presented
with a new graph of interest, can reliably and robustly predict the growth mechanism which
gave rise to that graph. Within the machine learning community, such
predictive, {\it supervised learning} techniques are differentiated from
descriptive, {\it unsupervised learning} techniques such as clustering.

We apply our method to the recently-published {\em Drosophila melanogaster} 
protein-protein interaction network \cite{giot} and find 
that a duplication-mutation-complementation mechanism 
\cite{flammini} best reproduces {\em Drosophila}'s network. 
The classification is robust against noise, even after 
random rewiring of 45\% of the network edges. To validate, 
we also show that beyond 80\% random rewiring the correct 
(Erd\"os-R\'enyi) classification is obtained.

\section{Methods}
\subsection{The data set}
We use a  protein-protein interaction map based on yeast 
two-hybrid screening ~\cite{giot}. Since  the data set is 
subject to numerous false positives, Giot {\em et al.}
assign a confidence score $p\in[0,1]$, measuring how likely 
the interaction occurs {\em in vivo}. 
In order to exclude unlikely interactions and focus on a 
core network which retains significant global features, 
we determine a confidence threshold $p^*$ based on percolation: 
measurements of the size of the components for all possible 
values of $p^*$ show that the two largest components are connected for 
$p^*=0.65$ (see supplemental material). Edges in the graph 
correspond to interactions for  which $p>p^*$. To reveal 
possible structural changes in {\em Drosophila} for 
less stringent thresholds, we also present results for $p^*=0.5$ 
as suggested in \cite{giot}.
We remove self-interactions from the network since none of the 
proposed mechanisms allow for them. After eliminating 
isolated vertices the resulting networks consist of 3359 (4625) 
vertices and 2795 (4683) edges for $p^*=0.65\; (0.5)$. 

\subsection{Network mechanisms}
We create 7000 graphs as training data, 1000 for each of 
seven different models drawn from the literature. Every graph is 
generated with the same number of edges and number of vertices 
as measured in {\em Drosophila}; all other existing parameters 
are sampled uniformly~\cite{droso_supp}.
The models manifest various simple network mechanisms, 
many of which explicitly intend to model protein interaction networks.

The duplication-mutation-complementation~\cite{flammini} (DMC) 
algorithm is inspired by an evolutionary model of the 
genome \cite{hughes,force} proposing that most of the 
duplicate genes observed today have been preserved by 
functional complementation. If either the gene or its 
copy loses one of its functions (edges), the other becomes 
essential in assuring the organism's survival. There 
is thus an increased preservation of duplicate genes induced 
by null mutations.
The algorithm features  a duplication step followed by 
mutations that preserve functional complementarity. At every time 
step one chooses a vertex $v$ at random. A twin vertex $v_{twin}$ 
is then introduced copying all of $v$'s  edges. For each edge 
of $v$, one deletes with probability $q_{del}$ either the original 
edge or its corresponding edge of $v_{twin}$. 
The twins themselves are conjoined with an independent 
probability $q_{con}$, representing an interaction of a 
protein  with its own copy. Note that no new edges are 
created by mutations. 
The DMC mechanism thus  assumes that the probability of 
creating new advantageous functions by random mutations is negligible.

A slightly different implementation of duplication-mutation 
is realized in~\cite{Sole} using random mutations (DMR). 
Possible interactions between twins are neglected. Instead, 
edges between $v_{twin}$ and the neighbors of $v$ can be 
removed with a probability $q_{del}$ and new edges can be 
created at random between $v_{twin}$ and any other vertices with a 
probability $q_{new}/N$, $N$ being the current total number 
of vertices. DMR thus emphasizes the creation of 
new advantageous functions by mutation.

Additionally, we create training data for linear preferential 
attachment (LPA)  networks \cite{barabasi,Price65} (growing 
graphs with a probability of attaching to previous vertices proportional 
to $k+a$, $a$ being a constant parameter, and $k$ 
the degree of the chosen vertex), random static networks 
(RDS) \cite{erdos} (also known as Erd\"{o}s-R\'enyi graphs; 
vertices are connected randomly), random growing 
networks (RDG) \cite{callaway}  (growing graphs where new edges 
are created randomly between existing vertices), aging 
vertex (AGV)  networks \cite{klemm} (growing graphs modeling 
citation networks, where the probability for new edges decreases 
with the age of the vertex), and small-world (SMW) 
networks \cite{watts} (interpolation between regular ring lattices 
and randomly connected graphs). For descriptions of the 
specific algorithms we refer the reader to the supplemental 
material.

\subsection{Subgraph census}
We quantify the topology of a network by exhaustive subgraph 
census \cite{wasserman} up to a given subgraph size;
note that we do {\em not} assume a specific network randomization nor 
test for statistical significance as in 
\cite{motifs:alon:nat,motifs:alon:science,motifs:hasty:nature,motifs:lee:science,motifs:barabasi:natgen,motifs:vespignani:natgen,motifs:alon:jmb,motifs:alon:pnas,matstat}, 
but we classify network mechanisms using the raw subgraph counts.
 Rather than choosing most important features a priori, we count 
all possible subgraphs up to a given cut-off,
which can be made either  in the number of vertices, number of 
edges, or the length of a given walk.
 To show insensitivity to this choice, we present results for 
two different cut-offs.
 We first count all subgraphs that can be constructed by a walk 
of length eight (148 non-isomorphic\footnote{Two graphs are isomorphic 
if there exists a relabeling of their vertices such 
that the two graphs are identical.} subgraphs); second, we 
consider all subgraphs up to a total number of 
seven edges (130 non-isomorphic subgraphs).
 Their counts are the input features for  our classifier.  
It is worth noting that the mean geodesic length (average shortest 
path between two vertices) of the {\em Drosophila} network's 
giant component is $11.6$ $(9.4)$ for $p^*=0.65\;(0.5)$. Walks of 
length eight are therefore able to traverse large parts of the network 
and can also reveal global structures.

\subsection{Learning algorithm}
Our classifier is a generalized decision tree called an 
{\em Alternating Decision Tree} (ADT) \cite{ADT} which uses the 
Adaboost \cite{adaboost} algorithm to learn the decision rules 
and associate weights to them.
Adaboost is a general discriminative learning algorithm 
proposed in 1997 by Freund and Schapire~\cite{adaboost97,adaboost}, 
and has since been successfully used in numerous and 
varied applications (e.g., in text 
categorization \cite{boostexter,boostinglit} and gene expression 
prediction \cite{geneclass}). It is equivalent to 
an additive logistic regression model \cite{friedman:logitboost}.

An example of an ADT is shown in Figure \ref{fig:tree}. 
A given network's subgraph counts determine paths in the 
tree dictated by inequalities specified by 
the {\em decision nodes} (rectangles).
 For each class, the ADT outputs a real-valued 
{\em prediction score}, which is the sum of all weights 
over all paths. The class with the highest score wins. 
The prediction score $y(c)$ for class $c$ is related to 
the probability $p(c)$ for the tested network to 
be in class $c$ by $p(c)=e^{2y(c)}/(1+e^{2y(c)})$~\cite{friedman:logitboost}. (The supplemental material gives details on the exact learning algorithm.)

An advantage of ADTs is that they do not assume a specific geometry 
of the input space; that is, features are not coordinates in a 
metric space (as in support vector machines or k-nearest-neighbors 
classifiers), and the classification is thus independent  
of normalization. The algorithm assumes neither independence nor 
dependence among subgraph counts. The features distinguish 
themselves solely by their individual abilities to discriminate 
different classes. 

\section{Results}
\begin{figure}[htb]
\begin{center}
\begin{tabular}{cc}
\includegraphics[height = 2.9in]{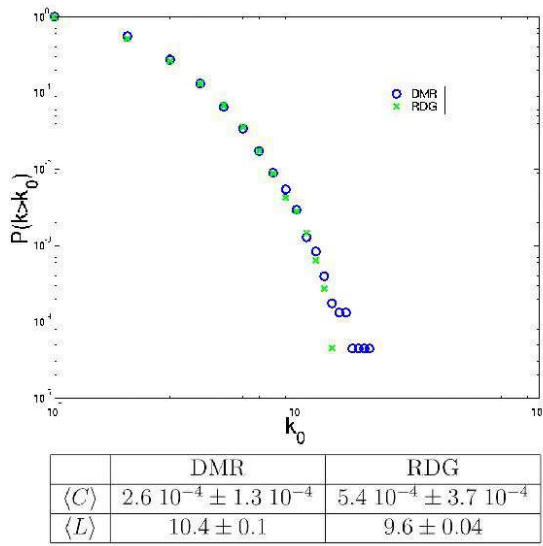}
&
\includegraphics[height = 2.9in]{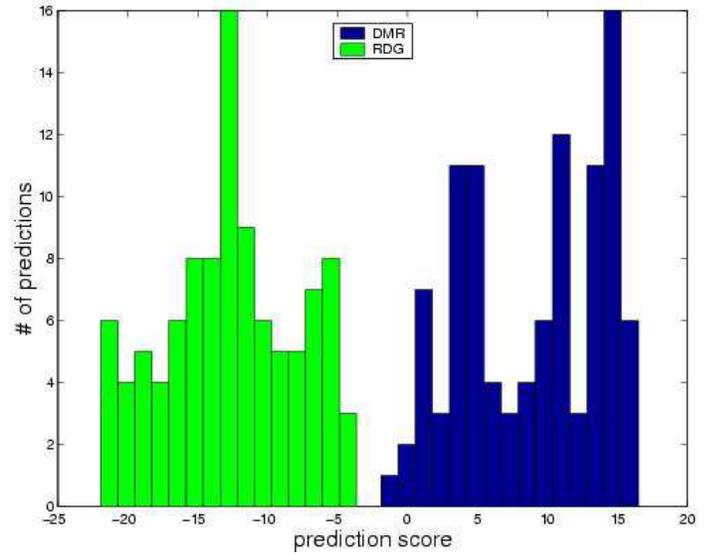}
\\
\bf{(a)} & \bf{(b)}\\
\end{tabular}
\end{center}
\caption{ {\bf Discriminating similar networks:}  Ten graphs of two different mechanisms exhibit similar average geodesic lengths and almost identical degree dstribution and clustering coefficients. {\bf (a)} cumulative degree distribution $p(k>k_0)$, average clustering coefficient $\langle C\rangle$ and average geodesic length $\langle L\rangle$, all quantities averaged over a set of ten graphs. {\bf (b)} prediction scores for all ten graphs and all five cross-validated~\cite{hastie} ADTs. The two sets of graphs can be perfectly separated by our classifier.} 
\label{fig:dmrgr}
\end{figure}

We perform cross-validation \cite{droso_supp,hastie} with multi-class 
ADTs, thus determining an empirical estimate of the 
generalization error, the probability of mislabeling an unseen test 
datum. The confusion matrix in Table \ref{tab:confusion} 
shows truth and prediction for the test sets. 
Five out of seven classes have nearly perfect 
prediction accuracy. Since AGV is constructed to 
be an interpolation between LPA and a ring lattice, the AGV, 
LPA and SMW mechanisms are equivalent in specific 
parameter regimes and correspondingly show a non-negligible 
overlap. Nevertheless, the overall prediction accuracy on 
the test sets still lies between 94.6\% and 95.8\% for 
different choices of $p^*$ and subgraph size cut-off. 
Note that preferential attachment is completely 
distinguishable from duplication-mutation despite the fact that
a duplication mechanism introduces an {\em effective} preferential
attachment~\cite{droso_supp,vazquez03}. Even models 
that are based on the same  fundamental mechanism, like 
duplication-mutation in DMC and DMR, are perfectly separable. 
Only small algorithmic changes in network mechanisms can thus 
give rise to easily detectable differences in substructures. 
Figure \ref{fig:dmrgr} confirms that although many of these 
models have similar degree distributions, clustering coefficients, 
or mean geodesic lengths, they have indeed distinguishable 
topologies.

\begin{figure}[htb]
\begin{center}
\resizebox{!}{5in}{\includegraphics{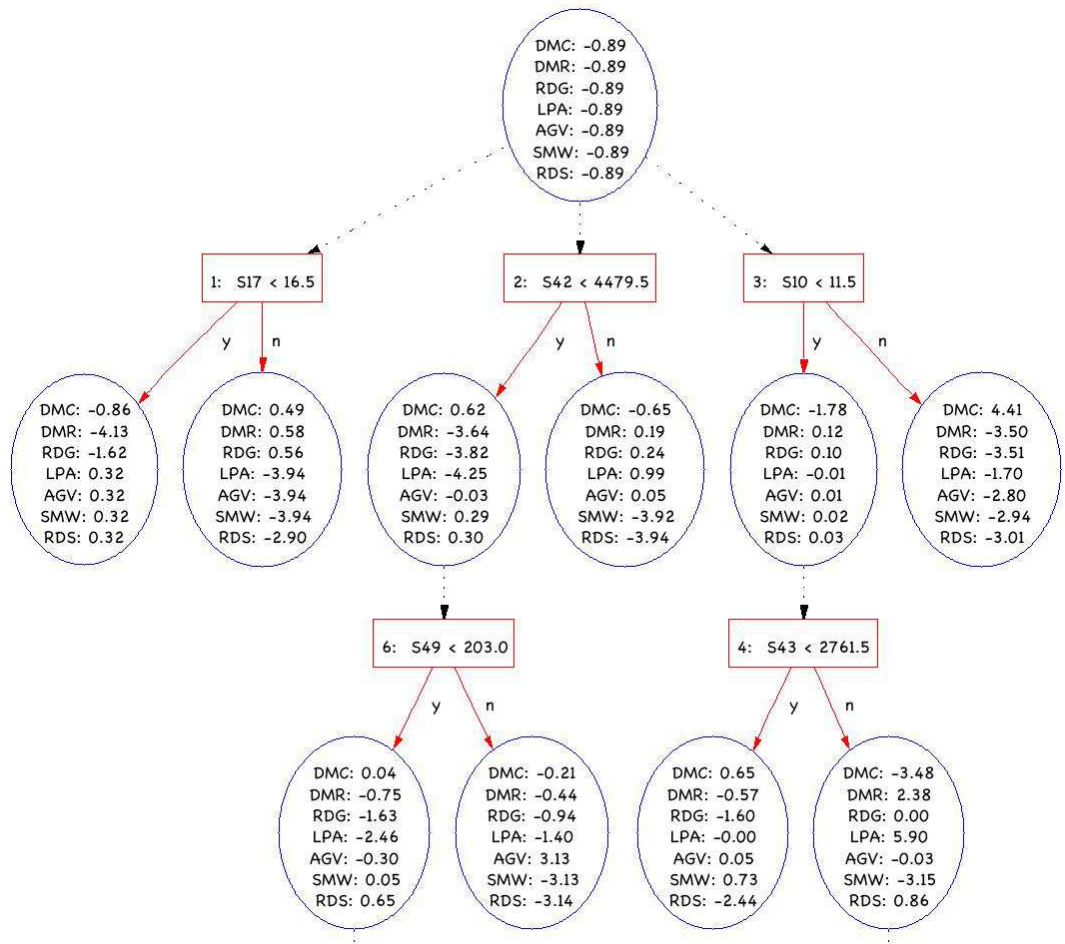}}
\end{center}
\caption{ {\bf Alternating decision tree:} The first few nodes of one of the trained ADTs are shown. At every boosting iteration one new decision node (rectangle) with its two prediction nodes (ovals) is introduced. Every test network follows several paths in the tree dictated by inequalities in the decision nodes (S\# refers to a specific subgraph count; see Figure \ref{fig:subgraphs}.). The final score is the sum of all prediction scores over all paths and the class with the highest prediction score wins.}
\label{fig:tree}
\end{figure}

Figure \ref{fig:tree} shows the first few decision nodes 
(out of 120) of a resulting ADT. 
The prediction scores
reveal that a high count of 3-cycles suggest a DMC network (node 3). 
The DMC  mechanism indeed facilitates the creation of many 
3-cycles by allowing two copies to attach to each other, 
thus creating 3-cycles with their common neighbors. 
In particular a few combinations are good predictors 
for some classes. For example, a low count in 3-cycles 
but a high count in 8-edge linear chains is a good predictor 
for LPA and DMR networks (nodes 3 and 4). Due to the 
sparseness of the networks the preferential attachment does 
not lead to a clustered structure. While LPA readily yields hubs, 
cycles are less probable. 
(More detailed ADTs can be viewed in the supplemental material.) 

Having built a classifier enjoying good prediction accuracy, 
we can now determine the network mechanism that 
best reproduces the {\em Drosophila} protein 
network (or in priniciple any network of same size) using the 
trained ADTs for classification. Table \ref{tab:pred} 
gives the prediction scores of the {\em Drosophila } network for 
each of the seven classes, averaged over folds. 

The duplication-mutation-complementation mechanism is the only 
class having a positive prediction score in every case. 
In particular for $p^*=0.65$ the DMC classification has a 
high score of 8.2 and 8.6. 
 Also, the comparatively small standard deviations 
over different folds indicate robustness of the classification 
against data subsampling.
While the high rankings of both duplication-mutation  classes confirm 
our  biological understanding of protein network  evolution,  
our findings strongly support an evolution restricted by 
functional complementarity over an evolution that creates and 
deletes functions at random. 
 
Interestingly for $p^*=0.65$ the RDG mechanism of random 
growth (edges are connected randomly between existing vertices) has 
a higher prediction score than the LPA or AGV growing graph mechanisms.
 Growth without any underlying mechanism other than chance 
therefore generates networks closer in topology to the core network
($p^*=0.65$) of {\em Drosophila} 
than growth governed by preferential attachment.
 We also emphasize that the small-world {\em character} 
of high clustering and short mean geodesic length, often 
attributed to biological networks \cite{giot,newyeast}, is 
not enough to conclude that the given network is close 
to the small-world {\em model} \cite{watts} (an interpolation between 
regular ring lattices and randomly connected graphs), as shown  here.
The classification for $p^*=0.5$ is less confident probably 
due to the additional noise present in the data when including low 
p-value (improbable) interactions, as we discuss below.

While not necessary for the classification itself, visualizing 
subgraph profiles can give a qualitative and more intuitive 
way of interpreting the classification result and a better 
understanding of the topological differences between {\em Drosophila} 
and each of the seven mechanisms. We plot in 
Figure \ref{fig:counts} their color-coded subgraph counts, 
averaged over all 1000 realizations of every model, for a representative 
subset of 50 subgraphs\footnote{We refer to the 
supplemental material for the whole set of 148 subgraphs}. We 
group together subgraphs (indicated by black lines) that exhibit 
the smallest absolute difference between the average subgraph 
count for the model, and for {\em Drosophila}.
For 60\% of the  subgraphs (S1-S30), {\em Drosophila}'s counts are closest to DMC's.  All of these subgraphs contain one or more cycles, including highly connected subgraphs such as K$_4$ (S1)\footnote{ a completely connected subgraph of four nodes}, and long linear chains ending in cycles (S16, S18, S22, S23, S25). 
DMC is the only mechanism that can give rise to the high occurrences of cycles measured in {\em Drosophila}.  Owing to the networks' sparseness cyclic structure is unlikely to be generated in  LPA, AGV, SMW, and RDS. The models LPA and AGV, however, are close to {\em Drosophila}'s topology according to subgraphs S44-S50 featuring open-ended chains and hubs, which occur frequently in both models as well as in {\em Drosophila}. 

\begin{figure}[hbt]
\begin{center}
\resizebox{!}{4.5in}{\includegraphics{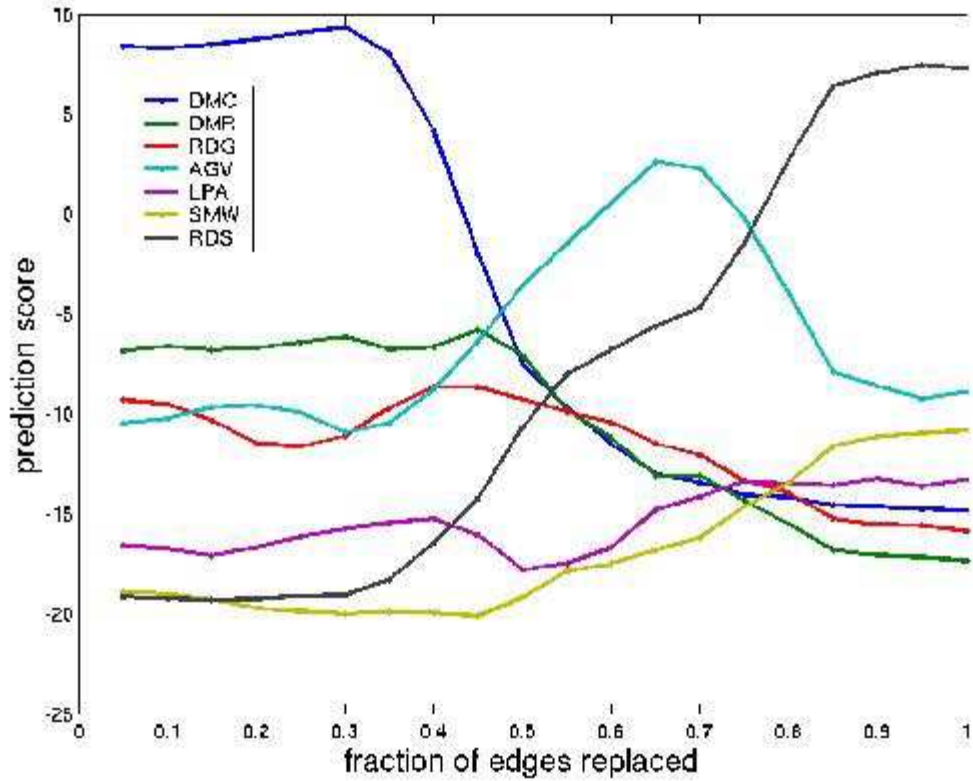}}
\end{center}
\caption{ {\bf Robustness against noise:} Edges in {\em Drosophila} are randomly replaced and the network is reclassified. Plotted are prediction scores for each of the seven classes as more and more edges are replaced. Every point is an average over 200 independent random replacements. For high noise levels (beyond 80\%) the network is classified  as  an Erd\"os-R\'enyi (RDS) graph. Also note that the confidence in the classification as a DMC network for low noise (less than 30\%) is even higher than in the classification as an RDS network for high noise. The prediction score $y(c)$ for class $c$ is related to the estimated probability $p(c)$ for the tested network to be in class $c$ by $p(c)=e^{2y(c)}/(1+e^{2y(c)})$~\cite{friedman:logitboost}.}
\label{fig:noise}
\end{figure}

Since yeast two-hybrid data is known to be susceptible 
to numerous errors~\cite{giot},  proposed inference methods 
are only reliable if they are  robust against noise. To 
confirm that our method shows this property, we classify the 
{\em Drosophila} network for various levels of artificially-introduced 
noise by replacing existing edges with random ones. 
Figure \ref{fig:noise} shows the prediction scores for all seven 
classes as functions of the fraction of edges replaced. 
As validation, the network is correctly classified as an RDS graph 
when all edges are randomized. About 30\% of {\em Drosophila}'s 
edges can be replaced without seeing any significant change in 
all seven prediction scores, and about 45\% can be replaced 
before {\em Drosophila} is no longer classified as a DMC network. 
At this point the prediction scores of DMC, DMR and AGV 
are very close, which is also observed for the 
prediction scores for $p^*=0.5$ (see Table \ref{tab:pred}), 
where they rank top three in this order. The results therefore 
suggest that the less confident classification for $p^*=0.5$ 
could be mainly due to the presence of more noise in the 
data after inclusion of low p-value edges.

We have presented a method to infer growth mechanisms for real 
networks. Advantageous properties include robustness both 
against noise and data subsampling, and the absence of any 
prior assumptions about which network features are important. Moreover, 
since the learning algorithm does not assume any 
relationships among features, the input space 
can be augmented with various features in addition to  subgraph counts.
We find that the {\em Drosophila} protein interaction network 
is confidently classified as a DMC network, a result which 
strongly supports ideas presented by Vazquez {\em et al.} \cite{flammini} 
and Force {\em et al.} \cite{force}  about the nature of genetic 
evolution. Recently, Wang {\em et al.} presented direct 
experimental evidence for a single DMC event 
in {\em Drosophila melanogaster} \cite{wang}. We 
anticipate that further use of machine learning techniques will answer 
a number of questions of interest in systems biology.

\baselineskip20pt

\begin{scilastnote}
\item It is a pleasure to acknowledge insightful discussions with Christina Leslie and Yoav Freund.
\end{scilastnote}

\newpage

\begin{table}
\begin{center}
\small{
\begin{tabular}{ll|ccccccc}
& & \multicolumn{7}{c}{{\sc Prediction}}\\
&  & DMR & DMC & AGV & LPA & SMW & RDS & RDG\\
\hline& DMR & 99.3\% &  0.0\% &  0.0\% &  0.0\% &  0.0\% &  0.1\% &  0.6\% \\
& DMC &  0.0\% & 99.7\% &  0.0\% &  0.0\% &  0.3\% &  0.0\% &  0.0\% \\
& AGV &  0.0\% &  0.1\% & 84.7\% & 13.5\% &  1.2\% &  0.5\% &  0.0\% \\
{\sc Truth}& LPA &  0.0\% &  0.0\% & 10.3\% & 89.6\% &  0.0\% &  0.0\% &  0.1\% \\
& SMW &  0.0\% &  0.0\% &  0.6\% &  0.0\% & 99.0\% &  0.4\% &  0.0\% \\
& RDS &  0.0\% &  0.0\% &  0.2\% &  0.0\% &  0.8\% & 99.0\% &  0.0\% \\
& RDG &  0.9\% &  0.0\% &  0.0\% &  0.1\% &  0.0\% &  0.0\% & 99.0\% \\
\end{tabular}
}
\end{center}
\caption{ {\bf Confusion matrix} for tested networks using five-fold cross-validation~\cite{hastie}. Entries $(i,j)$ show the probability of predicting class $j$ given that the true class is $i$. The training data is based on the size of the {\em Drosophila} protein network with a confidence treshold of $p^*=0.5$, the input features of the classifier being counts of all possible walks of length eight. The overall prediction accuracy is 95.8\%. Prediction errors among AGV, LPA and SMW networks are due to equivalence of the models in specific parameter regimes. }
\label{tab:confusion}
\end{table}

\begin{table}
\begin{center}
\begin{tabular}{|c||c|c||c|c||c|c||}
\hline
& \multicolumn{2}{c||}{\footnotesize 8-edge-walk subgraphs} & \multicolumn{2}{c||}{\footnotesize subgraphs with up to 7 edges} & \multicolumn{2}{c||}{\footnotesize 8-edge-walk subgraphs}\\
& \multicolumn{2}{c||}{\footnotesize $p^*=0.65$} & \multicolumn{2}{c||}{\footnotesize $p^*=0.65$} & \multicolumn{2}{c||}{\footnotesize $p^*=0.5$}\\
{\sc rank} & {\sc class} & {\sc score} & {\sc class} & {\sc score} & {\sc class} & {\sc score}\\
\hline
1 & DMC & $ 8.2 \pm 1.0    $& DMC & $ 8.6 \pm 1.1 $  & DMC & $ 0.8 \pm 2.9    $    \\
\hline                                               
2 & DMR & $ -6.8 \pm 0.9   $& DMR & $ -6.1 \pm 1.7 $ & DMR & $ -2.1 \pm 2.0    $   \\
\hline                                               
3 & RDG & $ -9.5 \pm 2.3   $& RDG & $ -9.3 \pm 1.6 $ & AGV & $ -3.1 \pm 2.2    $   \\
\hline                                               
4 & AGV & $ -10.6 \pm 4.2  $& AGV & $ -11.5 \pm 4.1 $& LPA & $ -10.1 \pm 3.1    $  \\
\hline                                               
5 & LPA & $ -16.5 \pm 3.4  $& LPA & $ -14.3 \pm 3.2 $& SMW & $ -20.6 \pm 1.9   $  \\
\hline                                               
6 & SMW & $ -18.9 \pm 0.7  $& SMW & $ -18.3 \pm 1.9 $& RDS & $ -22.3 \pm 1.7   $  \\
\hline                                               
7 & RDS & $ -19.1 \pm 2.3  $& RDS & $ -19.9 \pm 1.5 $& RDG & $ -22.5 \pm 4.7   $  \\
\hline
\end{tabular}
\end{center}
\caption{{\bf Prediction scores for the {\em Drosophila} protein network} for different confidence thresholds $p^*$ and different cut-offs in subgraph size. {\em Drosophila} is consistently classified as a DMC network, with an especially strong prediction for a confidence threshold of $p^*=0.65$  and independently of the cut-off in subgraph size.}
\label{tab:pred}
\end{table}

\newpage

\begin{figure}[hbt]
\begin{center}
\resizebox{!}{4.5in}{\includegraphics{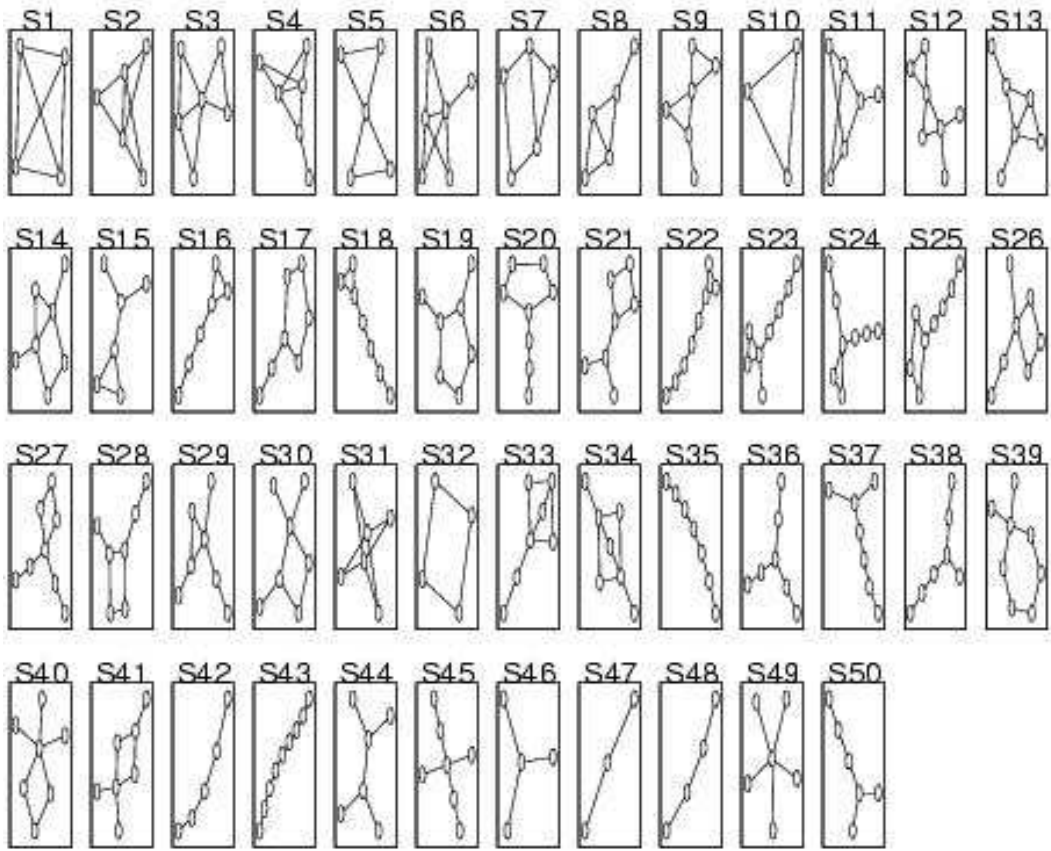}}
\end{center}
\caption{ {\bf Subgraphs associated with Figures \ref{fig:counts} and \ref{fig:tree}.} A representative subset of 50 subgraphs out of 148 is shown. }
\label{fig:subgraphs}
\end{figure}

\begin{figure}[hbt]
\begin{center}
\resizebox{!}{4.5in}{\includegraphics{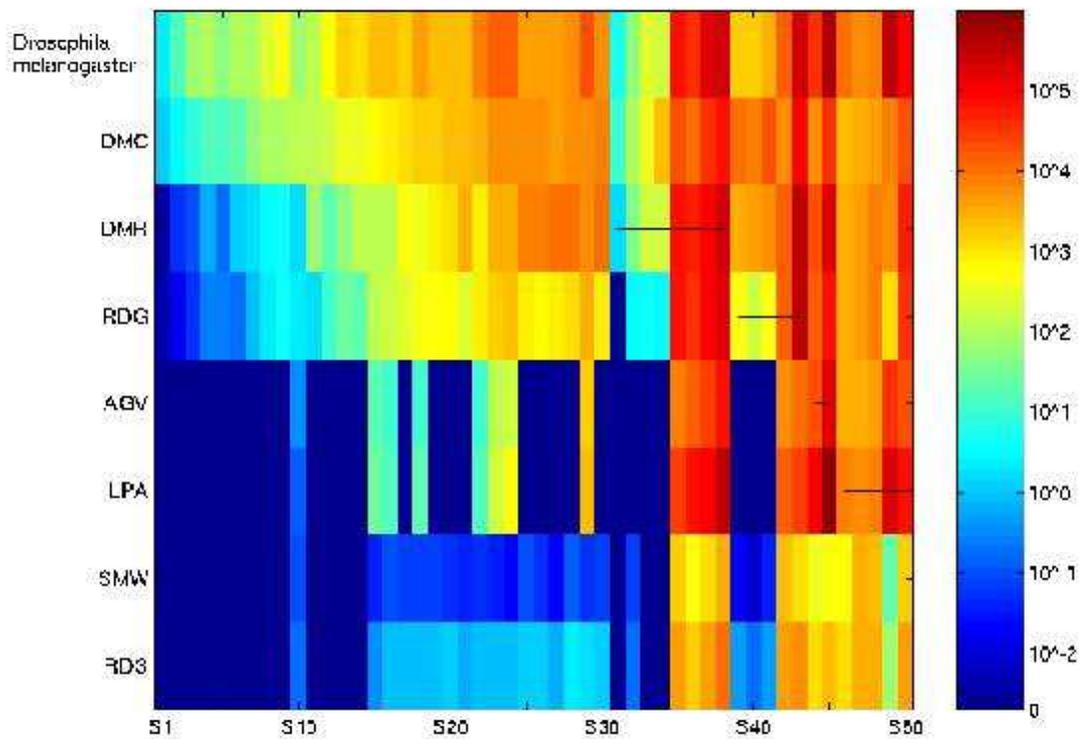}}
\end{center}
\caption{ {\bf Subgraph profiles.} The average subgraph count of the training data for every mechanism is shown for 50 representative subgraphs. The labels S1-S50 refer to Figure \ref{fig:subgraphs}. Black lines indicate that this model is closest to {\em Drosophila} based on the absolute difference between the subgraph counts. }

\label{fig:counts}
\end{figure}

\end{document}